\documentclass{article}
\usepackage{spconfa4,amsmath,amssymb,graphicx,xcolor}
\usepackage{array}
\usepackage{cite}
\title{ Graph Neural Network for Wide-Area 6DoF Ambisonic Rendering}

\name{Gioele Greco, Filippo Marri, Mirco Pezzoli, and Fabio Antonacci\thanks{}}
\address{Dipartimento di Elettronica, Informazione e Bioingegneria - Politecnico di Milano, Milan, Italy}

\newcommand{\vect}[1]{\boldsymbol{#1}}

\begin{document}
\ninept
\maketitle

\begin{abstract}
Wide-area 6DoF rendering from sparse Ambisonic recordings is commonly
addressed by interpolating parametric spatial descriptors rather than
the waveforms themselves. This preserves the fidelity of a nearby
measured signal and avoids phase artifacts caused by mixing distant
arrays. However, existing DirAC-based approaches rely on geometric
interpolation weights that do not depend on the acoustic scene observed
by the microphones. We propose a lightweight Graph Neural Network (GNN) that
refines only this interpolation stage. The known Ambisonic arrays and
the query position are represented as a graph whose nodes
contain local DirAC features and whose edges encode relative distance
and directional relations. The network predicts positive
query-conditioned weights over the measured arrays, while the target
energy, diffuseness, and direction of arrival are reconstructed by an
analytical DirAC layer. Experiments on unseen
rooms, unseen speech signals, and unseen array configurations show that
the proposed weighting improves energy reconstruction and
radiation-pattern similarity with respect to baseline interpolation.
\end{abstract}

\begin{keywords}
6DoF audio, Ambisonics, graph neural networks, DirAC, sound field reconstruction, scene adaptation
\end{keywords}

\section{Introduction}
\label{sec:introduction}

Immersive audio for virtual and augmented reality requires rendering
methods that enable six-degrees-of-freedom (6DoF) where the listener changes both orientation
and position. This requirement is also reflected in the MPEG-I Immersive Audio standard for interactive VR/AR audio rendering with 6DoF listener motion \cite{iso23090-4-2025}. While rotational rendering can be addressed with standard
Ambisonic decoding or binaural spatialization, translational motion is
more demanding because the sound field must be estimated at positions
where no microphone is available. This problem becomes particularly
difficult in wide-area scenarios, where the available Ambisonic arrays
are sparse and irregularly distributed.

A first class of methods approaches the problem as sound field
reconstruction. 
Wave-expansion, equivalent-source, and
kernel-based formulations estimate the pressure field from acoustic
measurements by imposing an explicit propagation model
\cite{pezzoli2025low,antonello2017rir,ueno2018krr,dame2025,Greco2025Parametric,6DOF_object,miotello2024physics}. 
These methods are physically interpretable and can provide accurate reconstructions
when their assumptions are satisfied. However, their usable bandwidth
and spatial extent are limited by the sampling geometry, by the number
of measurements, and by the validity of the underlying acoustic model.
These limitations are especially relevant for large listening regions and sparse microphone layouts, which remain central challenges in physics-informed and large-domain sound field estimation \cite{koyama2025piml,figueroaDuran2025largeDomains}.

A second direction is to reconstruct a representation that is useful for
rendering, rather than the full complex wave field. Parametric spatial
audio methods \cite{tylka2020fundamentals,politis2023wide,Greco2025Parametric,politis2017omdirac} follow this principle by describing the local sound field
through perceptually relevant time-frequency (TF) descriptors. Directional
Audio Coding (DirAC), for example, represents each TF bin
using energy, active intensity, direction of arrival, and diffuseness
\cite{pulkki2017dirac,politis2017omdirac}. These quantities do not
preserve the full pressure field, but they capture the directional and
direct-to-diffuse structure needed to control a binaural renderer. This
makes them attractive when the goal is perceptually plausible
navigation, rather than exact numerical reconstruction.

Wide-area 6DoF Ambisonic rendering builds directly on this idea. In the
method \cite{politis2023wide}, DirAC
parameters are estimated at multiple Ambisonic recording positions,
interpolated at the listener position, and then used to adapt the
binaural statistics of the closest recording. This avoids directly
mixing Ambisonic signals captured at distant positions, which would
introduce phase inconsistencies and comb-filtering artifacts. The method
is simple, lightweight, and effective enough to have been selected in
the MPEG-I 6DoF Immersive Audio Call for Proposals  \cite{politis2023wide,iso23090-4-2025}. Its strength is also
its main limitation: the interpolation weights are purely geometric and
therefore cannot adapt to the acoustic scene observed by the arrays.

This paper keeps the same parametric rendering structure and modifies
only the interpolation stage. Instead of replacing DirAC with a
black-box predictor, we use a GNN to estimate
scene-dependent interpolation weights from the measured DirAC
descriptors and the relative geometry of the sensor layout. This is a
natural formulation because the input is an unordered set of Ambisonic
arrays, the number of available arrays may vary, and the relevant
information is mostly relational. Graph networks provide a natural relational inductive bias for this type of data, while message-passing neural networks update node representations by aggregating pairwise interactions and are permutation-equivariant by construction \cite{battaglia2018relational,gilmer2017neural}.

The proposed model is deliberately constrained. The network does not
directly synthesize sound fields and does not regress arbitrary acoustic
parameters at the query position. It predicts positive weights over the
known arrays. The target energy, diffuseness, and direction of arrival
are then obtained through an analytical DirAC reconstruction layer. This
keeps the output interpretable and preserves the rendering assumptions
of the original wide-area DirAC pipeline, while allowing the
interpolation rule to depend on the measured scene.

The contribution of this work is threefold. First, we formulate
wide-area DirAC interpolation as a graph-based weighting problem over a
sparse Ambisonic array layout. Second, we introduce a coordinate-free
feature representation based on relative distances and directional
relations, avoiding dependence on a fixed coordinate system or sensor
ordering. Third, we show on unseen rooms, unseen speech signals, and
unseen array configurations that the learned weights improve energy
reconstruction and first-order Ambisonics (FOA) radiation-pattern similarity.
\section{Problem formulation}
\label{sec:problem}
Consider a distributed Ambisonic sensor network with $M$ known arrays
at positions
$\mathcal{S}=\{\vect{r}_1,\ldots,\vect{r}_M\}\subset\mathbb{R}^3$
and a query position $\vect{q}\in\mathbb{R}^3$ where the sound field
must be rendered. The $m$th array records a higher-order Ambisonics
(HOA) signal $\vect{x}_m(f,n)\in\mathbb{C}^{C}$, with
$C=(L+1)^2$ channels for spherical-harmonic (SH) order $L$, where
$f$ and $n$ denote the frequency-bin and short-time Fourier transform
(STFT) frame indices, respectively. Let
$\vect{x}^{\mathrm{FOA}}_m(f,n)\in\mathbb{C}^{4}$ denote the
FOA subset and let
\begin{equation}
\mathbf{C}^{\mathrm{FOA}}_m(f,n)
=
\mathbb{E}\!\left[
\vect{x}^{\mathrm{FOA}}_m(f,n)
\vect{x}^{\mathrm{FOA},H}_m(f,n)
\right]
\end{equation}
be its smoothed spatial covariance matrix. The standard DirAC
descriptors for each $(f,n)$ are then computed as
\begin{equation}
\begin{aligned}
E_m
&=
\frac{1}{2}
\operatorname{Tr}\!\left\{
\mathbf{C}^{\mathrm{FOA}}_m
\right\},
\quad
\vect{i}_m
=
\operatorname{Re}\!\left\{
\begin{bmatrix}
C^m_{1,4}\\
C^m_{1,2}\\
C^m_{1,3}
\end{bmatrix}
\right\},
\\
\psi_m
&=
1-
\frac{\lVert\vect{i}_m\rVert_2}
     {E_m+\varepsilon},
\quad
\vect{u}_m
=
\frac{\vect{i}_m}
     {\lVert\vect{i}_m\rVert_2+\varepsilon},
\end{aligned}
\label{eq:dirac_descriptors}
\end{equation}
where we omitted $(f,n)$ for readability,
$C^m_{i,j}
=[\mathbf{C}^{\mathrm{FOA}}_m(f,n)]_{i,j}$,
$\operatorname{Tr}(\cdot)$ denotes the matrix trace, and
$\operatorname{Re}\{\cdot\}$ extracts the real part.
Following
\cite{politis2023wide,pulkki2017dirac,politis2017omdirac},
we define the energy-normalized directional descriptor as
\begin{equation}
\vect{\eta}_m(f,n)
=
\frac{\vect{i}_m(f,n)}{E_m(f,n)+\varepsilon}
=
\bigl(1-\psi_m(f,n)\bigr)\vect{u}_m(f,n).
\label{eq:eta}
\end{equation}

The interpolation of the parameters is implemented in \cite{politis2023wide} using geometric weights $g_m(\vect{q})$ from the enclosing simplex and the estimate at the query position is given as
\begin{align}
E_{\mathrm{base}}(f,n;\vect{q}) &= \sum_{m=1}^{M} g_m(\vect{q}) E_m(f,n),
\label{eq:delaunayE}\\
\vect{\eta}_{\mathrm{base}}(f,n;\vect{q}) &= \sum_{m=1}^{M} g_m(\vect{q}) \vect{\eta}_m(f,n).
\label{eq:delaunayeta}
\end{align}
The corresponding $\psi_{\mathrm{base}}$ and $\vect{u}_{\mathrm{base}}$ follow from $\vect{\eta}_{\mathrm{base}}$. A nearby measured recording is then re-rendered so that its binaural statistics match these target parameters \cite{politis2023wide}. This avoids the comb-filtering artifacts of direct waveform mixing across distant arrays. The remaining problem is to replace $g_m(\vect{q})$ with scene-dependent weights that still preserve this analytical structure.

\section{Graph model}
\label{sec:graph}
We build a complete graph $\mathcal{G}=(\mathcal{V},\mathcal{E})$ with $M+1$ nodes. The first $M$ nodes correspond to the known arrays and the last node corresponds to the query position. A fully connected graph is used so that every node can exchange information with every other node, avoiding a hard neighborhood definition that would depend on a particular layout.

The node state is intentionally compact. For each measured array we use
\begin{equation}
\vect{h}^{(0)}_m(f,n)=\big[\log E_m(f,n),\; \psi_m(f,n),\; s(f),\; 1,\; 0\big],
\label{eq:nodefeat}
\end{equation}
where $s(f)$ is a normalized frequency tag. The graph is evaluated independently for each TF bin $(f,n)$. $s(f)$ makes the predicted weights
frequency-conditioned without exchanging information between
different frequency bins. This design keeps the model compact,
supports parallel frequency-wise inference, and is compatible with
the frequency-separable formulation of linear acoustic propagation. 

The query node uses $\vect{h}^{(0)}_q=[0,0,s(f),0,1]$. Absolute coordinates are excluded from the node features. This allows the network to operate on different sensor layouts, with different numbers of arrays, without learning a sensor-specific coordinate system or ordering.
Geometry and directionality enter through the edges. For nodes $i$ and $j$ we define
\begin{equation}
\vect{e}_{ij}=\big[d_{ij},\; \vect{u}_i^\top\hat{\vect{r}}_{ij},\; \vect{u}_j^\top\hat{\vect{r}}_{ij},\; \vect{u}_i^\top\vect{u}_j\big],
\label{eq:edgefeat}
\end{equation}
with $d_{ij}=\|\vect{r}_i-\vect{r}_j\|_2$ and $\hat{\vect{r}}_{ij}=(\vect{r}_j-\vect{r}_i)/(d_{ij}+\varepsilon)$. These terms describe sensor spacing, whether the intensity at node $i$ points toward node $j$, whether the reverse relation is also true, and whether the two local flows are aligned. The representation is coordinate-free, rigid translations and rigid rotations do not affect it. Thus, the model can adapt easily to new configurations.
The query node has no measured acoustic direction. We therefore set
$\vect{u}_q=\vect{0}$ when constructing edges involving the query.

With $\Lambda$ message-passing layers, the update at layer
$\ell\in\{1,\ldots,\Lambda\}$ is
\begin{align}
\vect{m}_{ij}^{(\ell)}
&=
\phi_{\mathrm{msg}}^{(\ell)}
\!\left(
[\vect{h}_i^{(\ell-1)},
 \vect{h}_j^{(\ell-1)},
 \vect{e}_{ij}]
\right),
\\
\vect{a}_i^{(\ell)}
&=
\sum_{j\neq i}\vect{m}_{ij}^{(\ell)},
\\
\vect{h}_i^{(\ell)}
&=
\phi_{\mathrm{upd}}^{(\ell)}
\!\left(
[\vect{h}_i^{(\ell-1)},
 \vect{a}_i^{(\ell)}]
\right).
\end{align}
The final node-wise head is shared across all measured nodes and is
conditioned on the final query embedding:
\begin{equation}
a_m
=
\phi_{\mathrm{out}}
\!\left(
[\vect{h}^{(\Lambda)}_m,
 \vect{h}^{(\Lambda)}_q,
 \vect{e}_{mq}]
\right),
\qquad
w_m=\operatorname{softplus}(a_m).
\label{eq:weights}
\end{equation}

The shared head makes the architecture permutation-equivariant and independent of $M$. Softplus gives positive but unnormalized weights, which is preferable to a softmax here because the model must be able to change absolute energy.
The output weights are converted to DirAC parameters by an analytical reconstruction layer,
\begin{align}
\hat{E} &= \sum_{m=1}^{M} w_m E_m,
%\label{eq:Ehat}
\quad \hat{\psi} = \frac{\sum_{m=1}^{M} w_m \psi_m}{\sum_{m=1}^{M} w_m},
\label{eq:psihat}\\
\hat{\vect{v}} &= \sum_{m=1}^{M} w_m E_m (1-\psi_m)\vect{u}_m,
\qquad
\hat{\vect{u}}=\frac{\hat{\vect{v}}}{\|\hat{\vect{v}}\|_2+\varepsilon}.
\label{eq:uhat}
\end{align}
The network therefore learns only how to combine measured nodes. The acoustic descriptors are still generated by closed-form equations.

\section{Training and test procedure}
\label{sec:training_test}

The network is trained offline on simulated Ambisonic room impulse
responses from the HARP dataset \cite{saini2025harp}. HARP provides
higher-order Ambisonic RIRs generated with the image-source method
over a large set of room geometries, absorption conditions, and
source-receiver configurations. In this work only the FOA subset is
used, since both the input features and the target DirAC descriptors
are computed from first-order components.
\subsection{Network configuration.}
In the reported experiments, the GNN uses $\Lambda=2$
message-passing layers. The functions
$\phi_{\mathrm{msg}}^{(\ell)}$ and
$\phi_{\mathrm{upd}}^{(\ell)}$ are implemented as multilayer
perceptrons (MLPs) operating on the concatenated message and update
features in the preceding equations. The node representations use a
hidden dimension of $48$. The output function
$\phi_{\mathrm{out}}$ is a shared node-wise MLP with scalar output,
which produces one log-weight $a_m$ for each measured node before the
softplus transformation. No node-specific parameters are used. The
complete model contains $27\,121$ trainable parameters. Since the graph
is fully-connected, the message-passing cost scales as
$\mathcal{O}(M^2)$ with the number of measured arrays.
\subsection{Training}
\label{subsec:training}
Training signals are obtained by convolving the FOA RIRs with dry
speech signals randomly selected from the TIMIT corpus
\cite{garofolo1993timit}. At each training iteration, the room,
the speech excerpts, and the spatial sampling of the receiver grid are
randomized. This prevents the model from learning a fixed waveform or
a fixed microphone layout, and instead encourages it to learn a
scene-dependent interpolation rule for the DirAC parameters.

All signals are sampled at $16$ kHz and transformed to the
TF domain using a Hann window of $512$ samples, a hop size
of $128$ samples, and an FFT size of $512$. The DirAC parameters
defined in Sec.~\ref{sec:problem} are then extracted from the FOA
spatial covariance matrices. Low-energy bins are discarded during
training by a simple energy threshold relative to the peak energy of
the current example, so that the loss is not dominated by bins where
the spatial descriptors are unreliable.

For each training sample, $M=5$ receiver positions are selected as the
known Ambisonic arrays and another receiver position is used as the
query. The known nodes provide the measured DirAC descriptors, while
the query node provides only its position through the edge features
defined in Sec.~\ref{sec:graph}. The network outputs positive,
unnormalized weights $w_m$ for the known nodes. These weights are then
converted into the reconstructed query descriptors by the analytical
layer in Eqs.~\eqref{eq:psihat}-\eqref{eq:uhat}. Therefore, supervision
is applied to the reconstructed acoustic quantities, not directly to
the weights.

The loss combines energy, direct-energy, and spatial
regularization terms:
\begin{equation}
    \mathcal{L}
    = \lambda_E \mathcal{L}_E
    + \lambda_{\rho} \mathcal{L}_{\rho}
    + \lambda_{\mathrm{reg}} \mathcal{L}_{\mathrm{reg}} .
    \label{eq:training_loss}
\end{equation}
The energy term is computed in logarithmic scale,
\begin{equation}
    \mathcal{L}_E =
    \mathbb{E}\left[
    \left(10\log_{10}(\hat{E}+\varepsilon)
    -10\log_{10}(E^{\star}+\varepsilon)\right)^2
    \right],
\end{equation}
where $(\cdot)^{\star}$ denotes the ground-truth query descriptor.
The direct-energy term compares the direct component proxy
$\rho=E(1-\psi)$,
\begin{equation}
    \mathcal{L}_{\rho} =
    \mathbb{E}\left[
    \left(\hat{E}(1-\hat{\psi})
    -E^{\star}(1-\psi^{\star})\right)^2
    \right].
\end{equation}

Finally, spatial smoothness is encouraged by perturbing the query
position with a small random displacement $\boldsymbol{\delta}$ and
penalizing changes in the predicted weights,
\begin{equation}
    \mathcal{L}_{\mathrm{reg}} =
    \mathbb{E}\left[
    \left\|
    \vect{w}(\vect{q}) -
    \vect{w}(\vect{q}+\boldsymbol{\delta})
    \right\|_2^2
    \right].
\end{equation}
In the reported configuration,
$\lambda_E=0.33$, $\lambda_{\rho}=0.33$, and
$\lambda_{\mathrm{reg}}=0.05$. The query and microphone positions are
also jittered by $0.1$ m during training to improve robustness to
small localization errors.
\subsection{Test}
The evaluation is carried out with a Monte-Carlo protocol on unseen
rooms and unseen speech material. Specifically, we test the trained
model on $25$ rooms that were never used during training. For each
room, dry speech excerpts not used in training are selected from TIMIT
and convolved with the corresponding FOA RIRs. For each room/audio
condition, three different five-array configurations are randomly drawn
from the available receiver grid, leading to $75$ independent test
configurations. In each configuration, the selected arrays are used as
known nodes and all remaining receiver positions are used as queries.
The same configurations are used for the baseline \cite{politis2023wide} and for the
GNN, so that the comparison isolates the effect of the interpolation
strategy.
\label{subsec:training}
\subsection{Evaluation Metrics}
\label{ssec:metrics}
To assess the reconstruction quality at the query $\vect{q}$, we compare the predicted parameters (denoted by $\hat{\cdot}$) against the ground truth ($\cdot^*$) over reliable TF bins. Define
\begin{equation}
\operatorname{NMSE}_x
=10\log_{10}
\frac{
\sum_{f,n}
\left(
\hat{x}(f,n)-x^{\star}(f,n)
\right)^2
}{
\sum_{f,n}
\left(
x^{\star}(f,n)
\right)^2
+\varepsilon
},
\qquad
x\in\{E,\psi\}.
\label{eq:nmse}
\end{equation}
where $\epsilon$ prevents division by zero. Specifically, $\operatorname{NMSE}_E$ measures the normalized
reconstruction error of the acoustic energy density over the reliable
TF bins, whereas $\operatorname{NMSE}_{\psi}$ measures the error in
diffuseness, i.e., in the direct-to-diffuse balance represented by the
DirAC model.

Converting parametric DirAC states back into waveform signals is an inherently lossy rendering step that can perceptually smear spatial cues. To assess the actual sound energy density map presented to the listener, we evaluate also the output signal. Let $D(\theta, \phi)$ be the directional energy map computed by steering the final reconstructed Ambisonic signals toward azimuth $\theta$ and elevation $\phi$. Let $\hat{D}_i$ and $D_i^{\star}$ denote the predicted and
ground-truth directional-energy values at grid direction $i$. Their
weighted structural correlation is
\begin{equation}
P
=
\frac{
\sum_i
\Phi_i
(\hat{D}_i-\hat{\mu})
(D_i^{\star}-\mu^{\star})
}{
\sqrt{
\sum_i
\Phi_i
(\hat{D}_i-\hat{\mu})^2
\sum_i
\Phi_i
(D_i^{\star}-\mu^{\star})^2
}
},
\label{eq:spherical_correlation}
\end{equation}
where $\hat{\mu}$ and $\mu^{\star}$ are the corresponding weighted
means and $\Phi_i \propto \cos(\phi_i)$ is a polar weight compensating for the grid distortion at the poles. This metric is bounded between $1$ (perfect structural match), $0$ (no correlation), and $-1$ (opposite spatial focus).

The reference method of \cite{politis2023wide}, used here as the
baseline, constructs a Delaunay triangulation of the array positions.
For each query position $\vect{q}$, the corresponding local simplex
determines the active arrays, and its barycentric coordinates define
the geometric interpolation weights
$g_m^{\mathrm{base}}(\vect{q})$. These weights satisfy
$g_m^{\mathrm{base}}(\vect{q})\geq 0,$
with
$\sum_{m=1}^{M}g_m^{\mathrm{base}}(\vect{q})=1,$
and zero weight assigned to arrays outside the active simplex. Thus, the
baseline estimates are obtained by \eqref{eq:delaunayE}-\eqref{eq:delaunayeta}.
Hereafter, the superscript
``$\mathrm{base}$'' identifies quantities obtained with the
Delaunay-based baseline.

For comparative analysis, the advantage of the proposed GNN over the
Delaunay baseline is defined as
\begin{equation}
\Delta\operatorname{NMSE}_x
=
\operatorname{NMSE}^{\mathrm{base}}_x
-
\operatorname{NMSE}^{\mathrm{gnn}}_x,
\qquad
\Delta P
=
P^{\mathrm{gnn}}-P^{\mathrm{base}}.
\end{equation}
Positive values, therefore, indicate an improvement of the GNN.
\section{Results}
\label{sec:results}

\begin{figure*}[t]
\centering
\includegraphics[width=\textwidth]{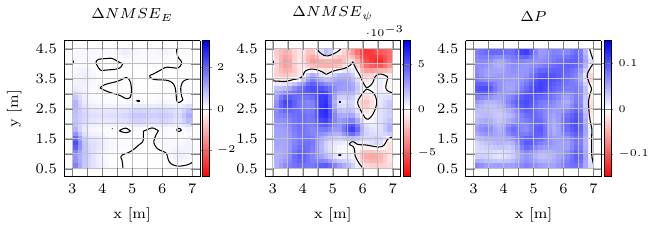}
\caption{Spatial distribution of the improvement obtained by the
proposed GNN with respect to baseline interpolation, averaged over the
Monte-Carlo test set. The panels report
$\Delta\mathrm{NMSE}_E$ for energy,
$\Delta\mathrm{NMSE}_{\psi}$ for diffuseness, and $\Delta P$ for
radiation-pattern similarity. Positive values indicate an improvement
of the GNN over baseline interpolation. Black contours mark the zero
level.}
\label{fig:spatial_maps}
\vspace{-2mm}
\end{figure*}

Fig.~\ref{fig:spatial_maps} reports the spatial distribution of the
performance difference between the proposed GNN and the baseline. The left panel shows the gain in energy reconstruction. The
GNN improves the energy NMSE over a large portion of the evaluated
region, with local gains reaching approximately
$\Delta\mathrm{NMSE}_E \simeq 3$dB. Negative regions are present but
remain spatially and numerically limited, with a maximum improvement of the baseline of $\simeq 0.1$dB. This indicates that the learned interpolation
weights are able to correct part of the mismatch between geometric
proximity and actual acoustic variation, which is one of the main
limitations of purely Delaunay-based interpolation \cite{politis2023wide}.

The center panel shows the corresponding result for diffuseness. In
this case, the GNN does not provide a clear improvement over the baseline, but it closely matches its behavior over most of the
domain. The median differences are on the order of $10^{-3}$, with
local variations of about $\pm 5\times10^{-3}$ and a standard deviation
mostly between $(3$ and $5)\times10^{-3}$, reaching approximately
$6\times10^{-3}$ only in localized regions. This is consistent with the
fact that the baseline is already particularly strong for
diffuseness. Unlike energy, diffuseness is bounded in
$[0,1]$ and varies more smoothly after the DirAC analysis, so a
geometry-based interpolation of nearby measurements is often sufficient
to obtain a reliable estimate. The relevant observation is therefore not
that the GNN outperforms the baseline on this descriptor, but that it
learns to reproduce a strong geometric estimator without degrading it
across unseen rooms, unseen speech signals, and unseen array
configurations.

The right panel in Fig.~\ref{fig:spatial_maps} reports the variation in
radiation-pattern similarity $\Delta P$. The GNN achieves a positive
gain over most of the domain, with local improvements approaching
$\Delta P \simeq 0.1$. This is a substantial increase: since Pearson
similarity is bounded, a gain of $0.1$ means that the predicted
directional pattern is not only rescaled, but is noticeably closer in
shape to the reference radiation pattern. The Monte-Carlo variability is
also consistent with this interpretation. The standard deviation of
$\Delta P$ is generally between $0.08$ and $0.12$, with localized values
up to approximately $0.16$ near boundary regions, where the randomly
selected array configurations can make the problem more sensitive to
the available spatial support. Thus, the gain is not uniform for every
room and array selection, but the median improvement remains positive
over almost the entire domain. In other words, the GNN better estimates
the angular structure that is later used by the DirAC rendering stage,
while retaining a clear advantage across unseen rooms, unseen speech
signals, and unseen array configurations.

\begin{figure}[tb]
\centering
\includegraphics[width=\columnwidth]{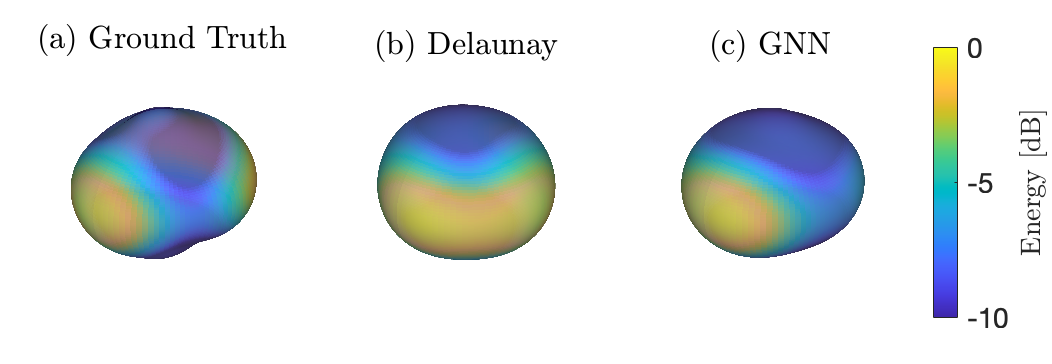}
\caption{Example of reconstructed FOA radiation pattern at one query
position and frequency band. The ground truth is compared with the baseline and the proposed GNN. Color represents normalized
energy in dB.}
\label{fig:radiation_example}
\vspace{-2mm}
\end{figure}

The qualitative example in Fig.~\ref{fig:radiation_example} confirms
the behavior observed in the spatial maps. The baseline
produces a smoother and more symmetric radiation pattern, losing part
of the asymmetric deformation present in the ground truth. The GNN
reconstruction better follows the dominant lobe and the overall
directional shape of the reference field. This is consistent with the
positive $\Delta P$ values in Fig.~\ref{fig:spatial_maps}: the learned
weights do not only reduce scalar descriptor errors, but also improve
the reconstruction of the directional FOA structure.

These results support the central assumption of the method,
replacing fixed geometric weights with scene-dependent graph weights
improves the interpolation of DirAC descriptors while preserving the
analytical reconstruction structure.
\section{Conclusion}
We proposed a graph-based refinement of wide-area DirAC interpolation
for 6DoF Ambisonic rendering. The method stays within the parametric
rendering pipeline, but replaces the fixed geometric interpolation rule
with a scene-dependent graph model. The sensor network is represented
explicitly, the feature design is coordinate-free, and the output
remains interpretable because the target DirAC parameters are
reconstructed analytically from positive node weights. This places the
method between classical parametric interpolation and data-driven sound
field reconstruction.

The current results show clear gains in energy reconstruction and
directional consistency, while diffuseness remains close to the
Delaunay baseline. This behavior is meaningful because the baseline is
already strong for diffuseness, whereas the learned model improves the
quantities for which geometry-only interpolation is more limited. The
small model size also makes the approach compatible with fast inference
and future scene-specific adaptation as new audio becomes
available.

Future work will extend the evaluation beyond simulated RIRs by testing
the method on real multi-point Ambisonic measurements. Perceptual
experiments, such as MUSHRA-style listening tests, will also be needed
to quantify whether the observed improvements in energy and
radiation-pattern similarity translate into audible improvements in
6DoF rendering quality. 
A further extension will investigate cross-frequency modeling. A
spectral encoder or autoencoder bottleneck could couple neighboring
frequency bins before graph aggregation, allowing the model to exploit
spectral continuity while retaining frequency-resolved DirAC
reconstruction.
Another promising direction is scene-adaptive
fine-tuning, where the pre-trained network is further optimized on a
short calibration recording from the target scene. This would preserve
the generalization provided by the synthetic training set while allowing
the graph weights to adapt to the specific room, source distribution,
and microphone layout used at test time.

\bibliographystyle{IEEEtran}
\bibliography{refs}

\end{document}